\documentclass[10pt, journal,twoside]{IEEEtran}
\usepackage{graphicx}
\usepackage{amsmath}
\usepackage{balance}
\usepackage{amssymb}
\usepackage{cases}
\usepackage[font=footnotesize,labelfont=bf]{caption}
\usepackage{caption}
\usepackage{color}
\usepackage{multirow} 
\usepackage{booktabs}
\usepackage[linesnumbered,ruled]{algorithm2e}
\usepackage[noend]{algpseudocode}
\usepackage[marginal]{footmisc}

\makeatletter
  \newcommand\figcaption{\def\@captype{figure}\caption}
  \newcommand\tabcaption{\def\@captype{table}\caption}
\makeatother
\usepackage{setspace}
\usepackage{bm}
\usepackage{setspace}
\usepackage{xcolor}
\usepackage{makecell}
\usepackage{booktabs}
\usepackage{multirow}
\usepackage{tabularx}
\usepackage{colortbl}
\usepackage{subcaption}
\usepackage{flushend} 

\begin{document}
\title{Private Data Leakage in Federated Human Activity Recognition for Wearable Healthcare Devices}
\author{Kongyang Chen, Dongping Zhang, Sijia Guan, Bing Mi, Jiaxing Shen, Guoqing Wang
\IEEEcompsocitemizethanks{
\IEEEcompsocthanksitem{Kongyang Chen and Dongping Zhang are with Institute of Artificial Intelligence, Guangzhou University, China. Kongyang Chen is also with Pazhou Lab, Guangzhou, China.}
\IEEEcompsocthanksitem{Sijia Guan is with School of Information Engineering, Wuhan University of Technology, Wuhan, China, and also with CASIC Space Engineering Development Co., Ltd., Beijing, China.}
\IEEEcompsocthanksitem{Bing Mi is with Pazhou Lab, Guangzhou, China.}
\IEEEcompsocthanksitem{Jiaxing Shen is with Department of Computing and Decision Sciences, Lingnan University, Hong Kong, China.}
\IEEEcompsocthanksitem{Guoqing Wang is with CSSC Huangpu Wenchong shipbuilding co., ltd., Guangzhou, China.}
}}

\IEEEtitleabstractindextext{
\begin{abstract}
Wearable data serves various health monitoring purposes, such as determining activity states based on user behavior and providing tailored exercise recommendations. However, the individual data perception and computational capabilities of wearable devices are limited, often necessitating the joint training of models across multiple devices. Federated Human Activity Recognition (HAR) presents a viable research avenue, allowing for global model training without the need to upload users' local activity data. Nonetheless, recent studies have revealed significant privacy concerns persisting within federated learning frameworks. To address this gap, we focus on investigating privacy leakage issues within federated user behavior recognition modeling across multiple wearable devices. Our proposed system entails a federated learning architecture comprising $N$ wearable device users and a parameter server, which may exhibit curiosity in extracting sensitive user information from model parameters. Consequently, we consider a membership inference attack based on a malicious server, leveraging differences in model generalization across client data. Experimentation conducted on five publicly available HAR datasets demonstrates an accuracy rate of 92\% for malicious server-based membership inference. Our study provides preliminary evidence of substantial privacy risks associated with federated training across multiple wearable devices, offering a novel research perspective within this domain.
\end{abstract}
\begin{IEEEkeywords}
Federated Learning, Human Activity Recognition, Membership Inference Attack, Privacy Leakage.
\end{IEEEkeywords}
}

\maketitle
\IEEEdisplaynontitleabstractindextext
\IEEEpeerreviewmaketitle

\section{Introduction}
In recent years, with technological advancements, there has been a continuous reduction in the size of chips and storage devices while their performance has steadily improved. This trend has propelled the enhancement of computational capabilities in wearable devices such as smartwatches, leading to their widespread applications across various domains \cite{Somani_Rogers_2024,JIANG2024107364}. Typically embedded with sensors like accelerometers and gyroscopes, these wearable devices collect real-time data and monitor users' health status. For instance, Human Activity Recognition (HAR) technology utilizes sensor data to identify human activities. Many smartphone manufacturers have introduced smart wristbands and watches with diverse functionalities, commonly equipped with motion recognition capabilities to automatically identify users' physical activities. The realization of these features relies on HAR models trained through Machine Learning to recognize user movements.

As the number of devices like smartwatches grows, so does the volume of data, making data privacy protection increasingly crucial. Traditionally, in HAR research, data collected by devices were often uploaded to servers for model training, with manufacturers integrating pre-trained models into devices before sale. However, this centralized data collection approach poses risks of data leakage. With numerous data breach incidents, data providers are increasingly averse to their privacy being compromised. Hence, since the proposition of federated learning by the Google team \cite{mcmahan2017communication, konecny2016federated}, researchers have begun adopting federated learning for HAR model training. Federated learning mitigates the risk of data leakage during data upload to servers, as individual participants only need to upload their trained models. Moreover, by utilizing federated learning to train HAR models, enterprises can distribute training tasks among many different participants. Under the premise of privacy protection, these participants are more inclined to use federated learning for model training as they do not wish to compromise privacy when uploading data to servers. For HAR applications, federated learning undoubtedly presents a viable solution, avoiding data leakage during participant data upload to servers, with original data stored locally and model training conducted locally.

However, several studies have indicated that even for pre-trained models, there remains a risk of privacy leakage. For example, Shokri et al. \cite{7958568} addressed the privacy risks inherent in machine learning as a service (MLaaS) by proposing Membership Inference Attack (MIA). MIA can infer whether an individual is included in the dataset used to train the target model, posing severe privacy risks to individuals. In many machine learning applications, the source of training data also carries privacy risks~\cite{chen2024fcl}. For instance, suppose there is a medical dataset used to train a model for predicting certain diseases. Attackers may use MIA to determine if a specific individual is in the dataset. For example, if the model predicts a person's medical condition \cite{jagannatha2021membership}, attackers may use MIA to determine if that person's data is in the training set, thereby obtaining their medical privacy information. In the field of speech recognition \cite{tseng2021membership}, attackers may use membership inference attacks to determine if someone's voice data was used to train the model. If attackers successfully determine that someone's data is in the training set, they may infer the person's voice characteristics and other privacy information. A report by the National Institute of Standards and Technology (NIST) specifically mentions that determining whether MIA infers an individual's inclusion in the dataset used to train the target model is a confidential act \cite{tabassi2019taxonomy}. Moreover, such privacy risks caused by MIA may lead commercial companies intending to provide MLaaS services to violate privacy regulations.

Therefore, this paper investigates the joint modeling process of multiple wearable devices aimed at user behavior recognition, with a focus on analyzing the privacy leakage issues associated with wearable data. Our analysis reveals significant differences in HAR data among different users. When employing joint modeling with multiple wearable devices, the prediction vector distributions of different users in each other's models differ substantially from their own models. This prompts an analysis of their privacy risks. Assuming the attacker as a curious server in federated learning possessing a certain amount of client data and partial knowledge of their sources, the objective is to distinguish data owned by but not attributed to any client. Hence, we explore the privacy risks present in the HAR model training process in federated learning. For the server, the training data from various clients are invisible, but during training, the models uploaded by individual clients are visible. By inputting known member and non-member data into the models uploaded by clients, the attacker obtains prediction vector distributions on the model and trains an attack model to distinguish data sources, known as Membership Inference Attack (MIA). MIA typically requires training data of the model, but in our approach, it only requires knowledge of which client the data belongs to, as the HAR models trained by clients usually generalize well to their own data but poorly to other clients' data. We conducted experiments on five publicly available HAR datasets under various conditions and observed the accuracy and recall rates of the attack model, demonstrating its ability to effectively differentiate remaining data with high accuracy.

The primary contributions of our study are as follows:
\begin{itemize}
  \item We consider the privacy leakage risks in user behavior recognition processes of wearable health devices. Our analysis reveals significant differences in HAR data among clients of multiple wearable devices, indicating potential data leakage issues. Particularly, we propose a method for analyzing user information leakage in federated HAR, aiming to provide a new research perspective for the field of federated learning in HAR. 
\item We conducted extensive experiments on five HAR datasets, demonstrating that attackers can exploit these differences for membership inference, achieving a maximum accuracy and recall of 92\%.
\end{itemize}

The remainder of the paper is structured as follows: Section~\ref{sec:motivation} describes our research motivation. Section~\ref{sec:method} presents our research methodology. Section~\ref{sec:experiment} provides a detailed explanation of the experimental results. Section~\ref{sec:related} elaborates on related work. Finally, Section~\ref{sec:conclusion} concludes the paper.

\section{Motivation}
\label{sec:motivation}
\textit{Human Activity Recognition (HAR): }
HAR stands as a significant research domain within artificial intelligence and signal processing, aiming to utilize data collected from various sensors such as accelerometers and gyroscopes to identify and classify human activities~\cite{chen2024FUHAR}. HAR finds applications in diverse fields, including health monitoring, fall detection, and smart homes~\cite{rosaline2023enhancing,yadav2022arfdnet,bianchi2019iot, yadav2022arfdnet}. Typically, HAR data is directly collected from sensors affixed to the human body, thereby increasing the potential risk of privacy exposure once malicious actors gain access to the data. Traditional HAR methods often rely on centralized data from different users or devices to train global models. However, this centralized approach raises concerns regarding privacy, security, and scalability \cite{zhou2014learning}.

In the traditional machine learning paradigm, collected data is usually uploaded to servers for model training. However, this method of uploading datasets to servers poses numerous potential risks of data leakage. For instance, incidents of data privacy breaches, such as those observed in companies like AOL~\cite{barbaro2006face}, have made both enterprises and individuals reluctant to easily upload their data to servers for model training, leading to the emergence of the so-called \textit{data islands} issue. The introduction of federated learning technology partially addresses this problem. Federated learning involves training models on client devices using local data and then uploading model parameters, losses, gradients, and other information to servers, which then aggregate these parameters to obtain a global model. This training approach allows client devices to avoid uploading local data to servers, effectively mitigating potential risks of data leakage during the upload process. With the advent of federated learning technology, researchers have begun exploring how to utilize federated learning to train HAR models, addressing issues such as data privacy leaks and data islands.

\begin{figure*}[t]
  \centering
  \begin{minipage}[t]{0.42\textwidth}
    \centering
    \includegraphics[width=\textwidth]{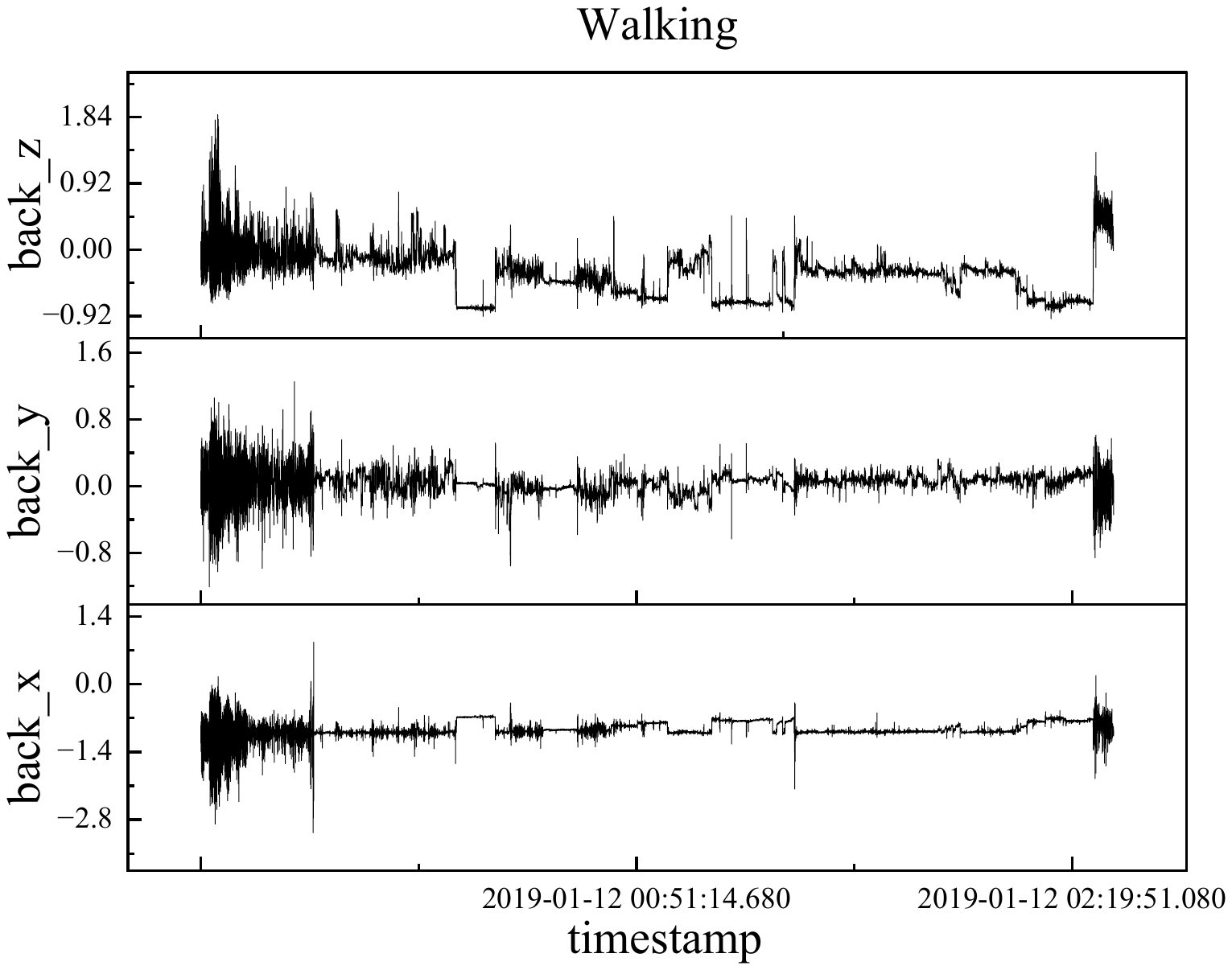}
  \end{minipage}
  \begin{minipage}[t]{0.42\textwidth}
    \centering
    \includegraphics[width=\textwidth]{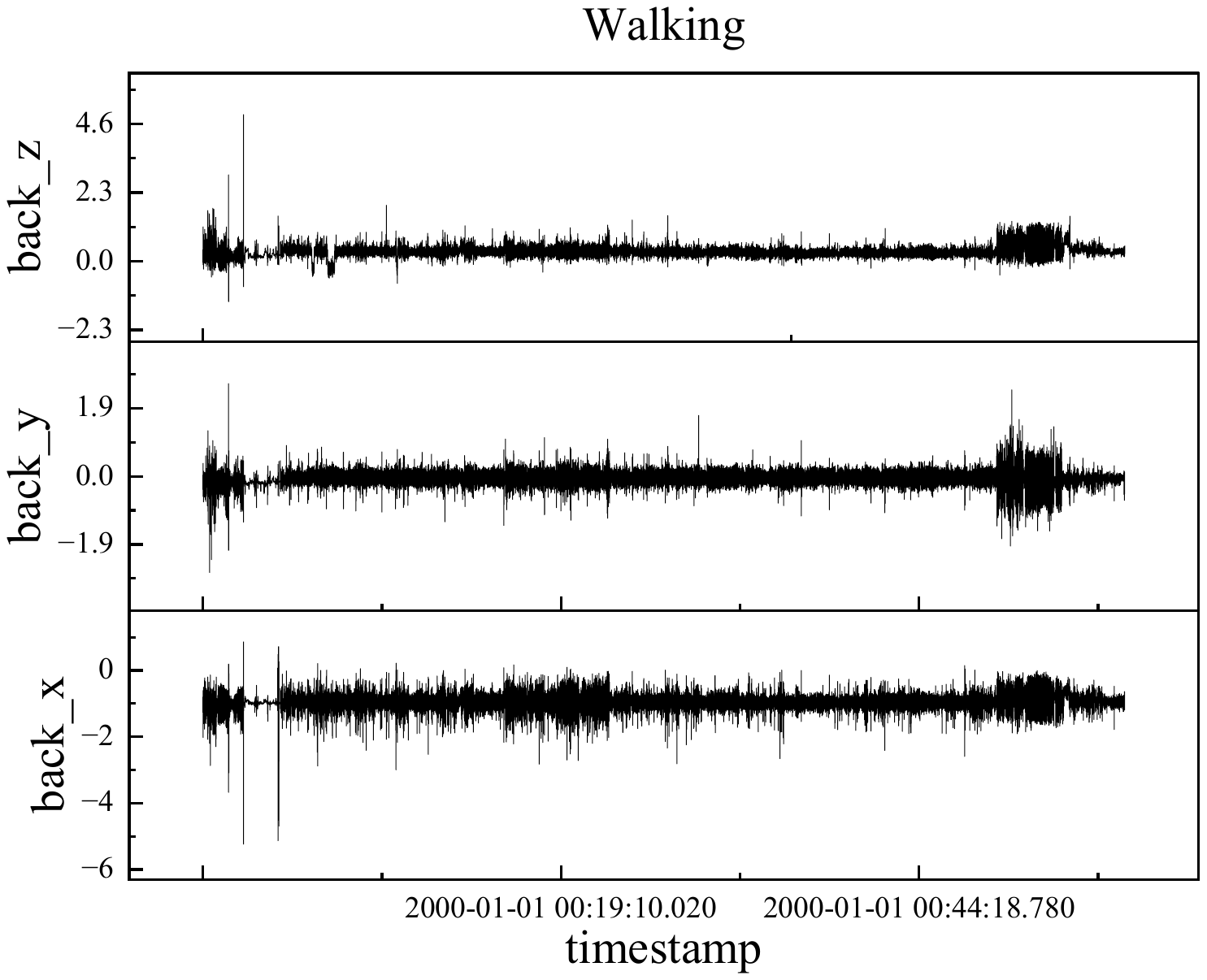}
  \end{minipage}
  \caption{The walking data of two subjects in the HARTH dataset.}
\end{figure*}

\begin{figure}[t]
  \centering
  \includegraphics[width=0.7\linewidth]{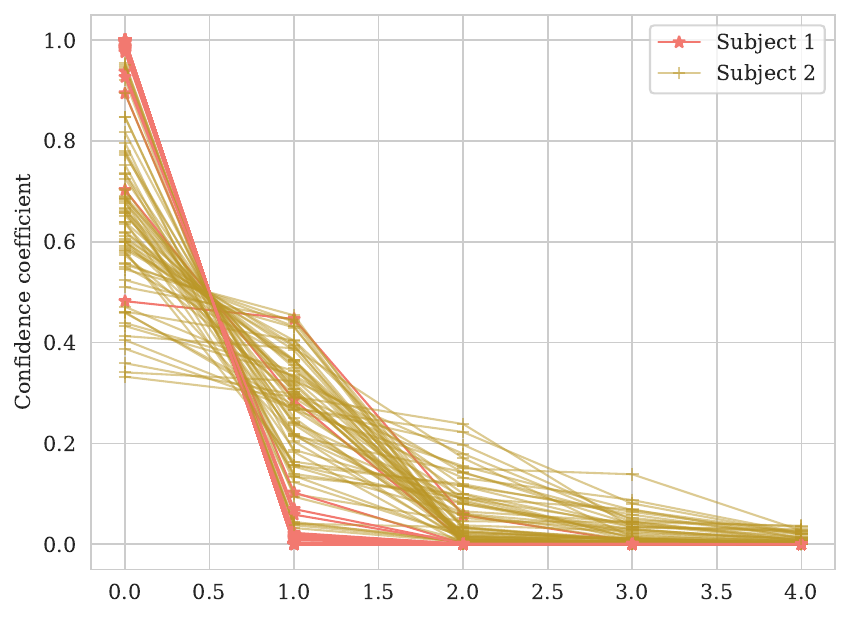}
    \caption{The model trained from subject 1 in the HARTH dataset.}
  \label{fig:predict_vector}
\end{figure}

\textit{Private Data Leakage In HAR: }
In many cases, federated learning addresses non-iid problems. However, there may exist varying privacy risks among the client devices involved in model training. In real-world application scenarios, HAR data collected from different individuals may exhibit differences due to individual habits, resulting in variations in the magnitude and posture of actions performed when executing the same activity. For instance, as depicted in Figure \ref{fig:harth_c1_c2_walking}, the sensor data signals on the back in the \textit{walking} activity data from subject 1 and subject 2 in the HARTH dataset exhibit significant differences despite representing the same activity. Consequently, due to the disparities in HAR data from different individuals, when training models using data from a single subject, the features learned by the trained model originate predominantly from that individual. As illustrated in Figure \ref{fig:predict_vector}, when training the model using only subject 1's data, the highest confidence in prediction vectors is consistently close to 1, and the distribution of data is similar. When subject 2's data is input into the model trained with subject 1's data, the highest confidence in the prediction vectors is only occasionally close to 1, with significant differences in the distribution of prediction vectors compared to those obtained from subject 1.

For attackers, the aforementioned disparities in output distributions of HAR data between different subjects on the model serve as valuable aids in determining the data source. Specifically, considering the scenario of federated learning involving two clients, each possessing HAR data from different individuals, attackers, upon acquiring HAR data, would attempt to input this data into models trained by clients to obtain corresponding prediction vectors and observe the distribution of these data. Leveraging the knowledge obtained from the aforementioned process, attackers can identify data most likely to belong to the client that uploaded the model with prediction vectors exhibiting the highest confidence close to 1. Subsequently, attackers may gather additional privacy information through the application scenarios of the trained model or the hardware information of the client device itself.

\section{Wearable Private Data Leakage in Federated Human Activity Recognition}
\label{sec:method}
This section introduces method for Federated HAR, discusses the issue of data leakage, and proposes a data leakage method based on member inference attacks.
\subsection{Federated HAR}

Different from the centralized HAR model training approach, the centralized training method collects data from smart devices worn by individuals and uploads it to a server for centralized model training. In contrast, federated learning enables these smart devices to act as clients. Suppose there are $n$ smart devices collecting data for model training. They are represented as clients in federated learning, denoted as $C=\{C_i\}_{i=0}^{n-1}$. These smart devices collectively train a global model $M_s$. Each client has its own HAR training dataset $D_i$ and a local model $M_i$ with the same structure as the global model. 
\begin{figure*}[!t]
  \centering
  \includegraphics[width=0.7\textwidth]{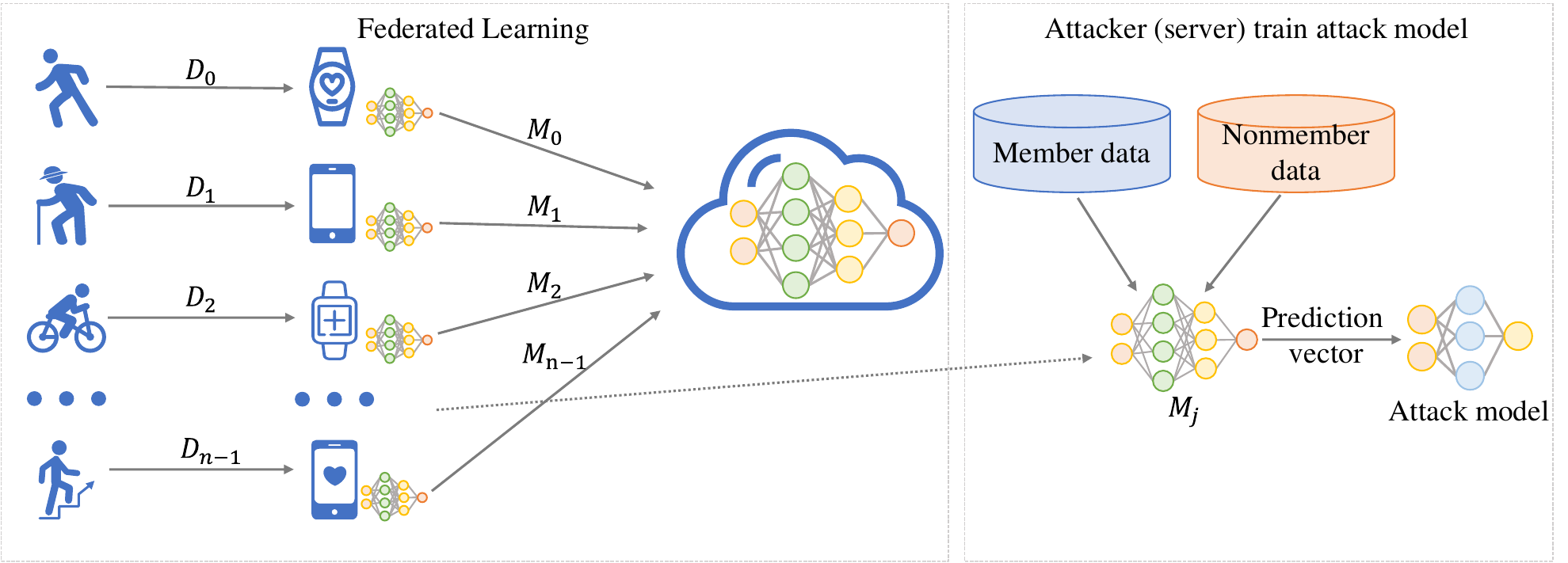}
  \caption{The general framework of data leakage in Federated HAR.}
  \label{fig:overview}
\end{figure*}
At the beginning of each training round, the server distributes the global model parameters $W$ to each client. Upon receiving the global model parameters, the client updates its local model to match the global model. Each training round randomly selects a subset of clients, denoted as $S$, for training. The selected clients use their local data to minimize the loss function $\mathcal{L}(w_{i}x_{i},y_{i})$ via stochastic gradient descent (SGD) to optimize the model parameters $w_i$, resulting in $M_{i}^{e}$, where $e$ represents the current training round. After local training, the clients upload their latest model  $M_{i}^{e}$ to the server, which aggregates these uploaded models $W = \frac1{|S|}\sum_{i\in S} w_i$ to obtain a new global model. This process repeats until convergence. 

\subsection{Wearable Private Data Leakage in Federated HAR}
With the proliferation of the Internet of Things (IoT), wearable devices such as smartwatches are equipped with various sensors such as accelerometers and gyroscopes, which can be used for activity detection. The combination of these sensors enables activity recognition through smartwatches, and HAR is increasingly used for identifying activities in daily life ~\cite{Dentamaro2024human}. The emergence of federated learning allows application developers or research institutions interested in training HAR models to collaborate with different HAR data owners, treating them as participating clients in federated learning to jointly train models. However, there are also malicious attackers who attempt to explore the privacy information of participating clients by training models. Attackers, with access to their own data, can invite data owners to train HAR models and simultaneously train attack models during the process. They use the attack models to determine the origin of subsequently obtained data.

In a HAR application trained via federated learning by $n$ clients $C={C_i|}_{i=0}^{n-1}$, there exists a curious server attacker whose goal is to determine whether data originates from one of the participating clients, $C_j$, upon acquiring the data, in order to obtain more privacy information.

To facilitate experimentation, we simplify the attacker's objective. Assuming the attacker obtains data belonging to client $C_j$ from a public database or other sources in the real world, denoted as $D_{member}$, along with some data belonging to other clients, $D_{nonmember}$. Additionally, the attacker possesses some data from clients whose origins cannot be distinguished, denoted as $D_{mix}$. The attacker's objective is to identify data belonging to client $C_j$ from this mixed dataset.

As the attacker cannot directly access client data in federated learning, they cannot compare it directly. However, as an intelligent and curious server, the attacker knows that the data of client $C_t$ will never participate in the training of client $C_j$'s model ($t \neq j$). Therefore, the attacker can exploit differences between HAR data from different clients by inputting known-source data into the model trained by client $C_j$ to obtain different prediction vectors $\hat{p}(y | x)$. These vectors serve as training data for an attack model $\mathcal{M}$. Specifically, after the target client uploads the model, the attacker saves it and utilizes their known data from the target client $D_{member}$ and data from non-target clients $D_{nonmember}$ to make predictions using the uploaded model. The attacker then uses these predictions to train attack model $\mathcal{M}$. Moreover, throughout the federated learning process, the attacker continuously improves the attack model's performance by saving models uploaded after each training round of the target client. Figure \ref{fig:overview} illustrates the process of the attacker training the attack model. After completing the training of the attack model, the attacker inputs data $D_{mix}$ into $\mathcal{M}$ for prediction. If the data belongs to client $C_j$, the output is 1; otherwise, it is 0.

\subsection{Membership Inference Attacks for Private Data Leakage}
In this process, we assume that the attacker possesses known member data $D_{\text{member}}$ and non-member data $D_{\text{nonmember}}$, as well as mixed data $D_{\text{mix}}$ that the attacker aims to differentiate. These are defined as follows:

\begin{equation}
    D_{member}=\left\{(x_i,y_i)\right\}_{i=1}^{N^m}.
\end{equation}
\begin{equation}
    D_{nonmember}=\left\{(x_i,y_i)\right\}_{i=1}^{N^n}.
\end{equation}
\begin{equation}
    D_{mix}=\left\{(x_i,y_i)\right\}_{i=1}^{N^x}.
\end{equation}

Let $C_j$ be the target client the attacker aims to exploit. Therefore, $D_{\text{member}} \subset D_j$ and $D_{\text{nonmember}} \subset D_{\text{other}}$, where $D_{\text{other}}$ may consist of mixed data from one or multiple clients. $D_{\text{mix}}$ contains data from both $C_j$ and other clients, with the number of client sources denoted as $k$. The attacker's objective is to distinguish data belonging to $C_j$. Thus, during each round of global model training, if $C_j$ is selected for training, the attacker saves the uploaded model $M_{j}^{e}$. The attacker inputs data into the model to obtain prediction vectors $\hat{p}(y\mid x)$, defined as follows:

\begin{equation}
   \hat{p}(y\mid x) =  sort(softmax(M_{j}^{e}(x))).
\end{equation}

\begin{figure}[t]
  \centering
  \includegraphics[width=0.42\textwidth]{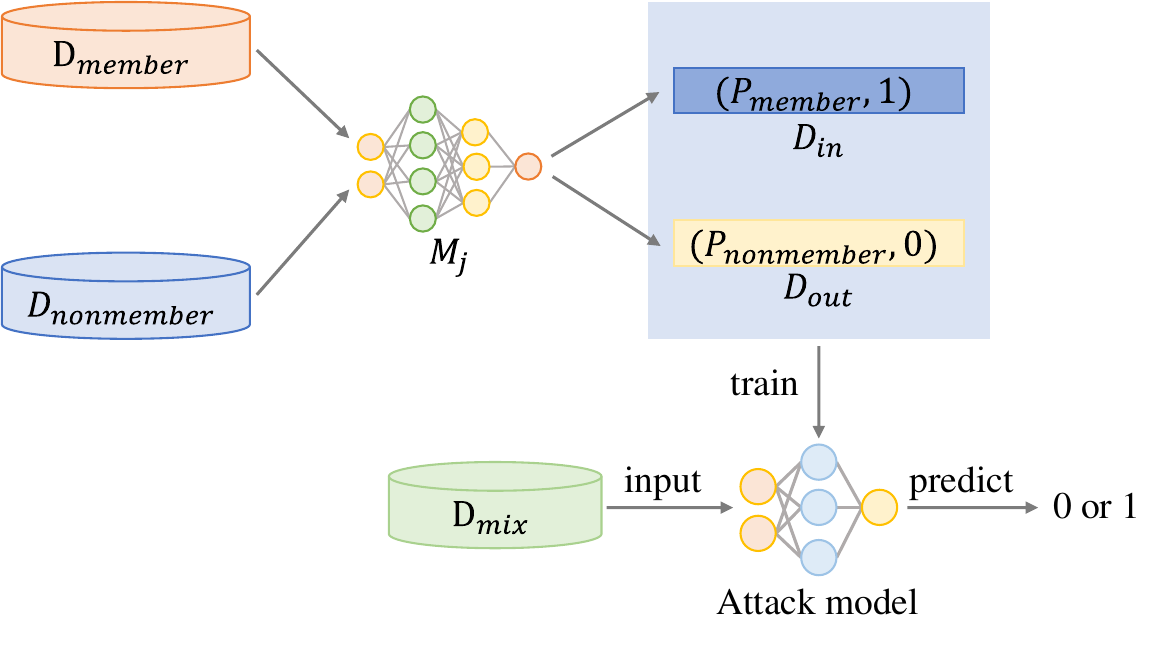}
  \caption{The process of the attacker training and utilizing the attack model to differentiate data after saving the uploaded model from the target client.}
  \label{fig:mia_train_predict}
\end{figure}

\begin{algorithm}[t]
    \caption{Attack model training algorithm.}
    \small
    \label{alg1}
    \KwData{Training round $Epochs$;Member data $D_{member}$;Nonmember data $D_{nonmember}$;Target Client ID $j$;}
    \KwResult{attack model $\mathcal{M}$;}
    \SetAlgoNlRelativeSize{}
    \SetNlSty{normalfont}{}{:}
    \For{$e \leftarrow 1$~\KwTo~Epochs} {
        $S \leftarrow$ randomly select $m$ clients for training;\\
        \If{$e == 1$} { 
            add $C_j$ to $S$;
        }
        \For{$C_i \in S$} {
            $w_i \leftarrow W$;\\
            $w_i \leftarrow$ Local SGD; 
        }
        \If{$C_j \in S$} {
            sever obtained the model $M_{j}^{e}$;\\
            $\hat{p}(y\mid x) = sort(softmax(M_{j}^{e}(x)))$;\\
            $D_{in} = \left\{(\hat{p}(y_i\mid x_i),1)\mid x_i\in D_{member}\right\}_{i=1}^{N^m}$; \\ 
            $D_{out} = \left\{(\hat{p}(y_i\mid x_i),0)\mid x_i\in D_{nonmember}\right\}_{i=1}^{N^n}$;\\
            \textbf{train attack model $\mathcal{M}$:}\\
            $\mathcal{M}(\hat{p}(y\mid x);\theta) = \mathcal{F}(\{D_{in}, D_{out}\})$;
        }
        $W = \frac1{|S|}\sum_{i\in S} w_i$;\\
    }
\end{algorithm}

We denote the sets of prediction vectors obtained from $D_{\text{member}}$ and $D_{\text{nonmember}}$ on the model as $P_{\text{member}}$ and $P_{\text{nonmember}}$, respectively. The data in $P_{\text{member}}$ is labeled as 1, while the data in $P_{\text{nonmember}}$ is labeled as 0, resulting in the training datasets $D_{\text{in}}$ and $D_{\text{out}}$ for the attack model $mathcal{M}$, trained using learning algorithm $\mathcal{F}$.

\begin{equation}
    \begin{aligned}
        D_{in}= \left\{(\hat{p}(y_i\mid x_i),1)\mid x_i\in D_{member}\right\}_{i=1}^{N^m}, \\ 
        D_{out}= \left\{(\hat{p}(y_i\mid x_i),0)\mid x_i\in D_{nonmember}\right\}_{i=1}^{N^n}.
    \end{aligned}
\end{equation}

The overall training and prediction process of the attack model $\mathcal{M}$ is illustrated in Figure \ref{fig:mia_train_predict}, where an output of 1 indicates the model predicts the record belongs to $C_j$, while an output of 0 indicates the record belongs to another client.

\begin{equation}
    \mathcal{M}(\hat{p}(y\mid x);\theta)=\mathcal{F}(\hat{p}(y\mid x),y).
\end{equation}
\begin{equation}
    \mathcal{M}(\hat{p}(y\mid x);\theta)=
    \begin{cases}
        1&\mathrm{if~}x\in C_j,\\
        0&\mathrm{if~}x\notin C_j.
    \end{cases}
\end{equation}

Here, $\theta$ represents the parameters of model $\mathcal{M}$, and the pseudo-code for the algorithm is presented in Algorithm~\ref{alg1}.

\section{Experiment Results}
\label{sec:experiment}
In our study, we conducted extensive experimental validation across five publicly available HAR datasets. These experiments encompassed diverse model selections, training datasets, and training methods to ensure the comprehensiveness and reliability of our research. Furthermore, we delved beyond mere data collection and simple experimental validation, undertaking an in-depth exploration of the impact of federated learning attacks. We also employed common MIA defense mechanisms, such as Dropout and $L_2$ regularization, to enhance the robustness of the models. These measures not only enriched the scope of our study but also ensured thorough consideration of the experimental results.

\subsection{HAR datasets}
This subsection provides detailed descriptions of the datasets used in the experiments, including their sources, features, and scales. 

In our experiments, we validated our approaches using five publicly available HAR datasets. These datasets encompass HAR data collected under various environmental conditions, sensor types, and age groups, making them naturally suitable for FL experimental settings, with each subject in the dataset treated as a participating client. Additionally, the varying sizes of these datasets allow for a broader assessment of the privacy risks posed by attacks.

\begin{table*}[t]
  \centering
  \caption{The study encompasses five datasets gathered through various sensing devices in diverse environmental settings, each dataset varying in size.}
  \label{tab:datasets}
  \begin{tabular}{cccc}
    \toprule
    \textbf{Dataset} & \textbf{\makecell{(\# Subject,   \\~~~~~\# Activities)}} & \textbf{Task} & \textbf{Sensor}\\
    \midrule
    \textbf{UCI HAR} & (30, 6)                        &
    \makecell{Walking/Laying \\ Sitting/Standing\\ Upstairs/Downstairs} & \makecell{Accelerometer \\ Gyroscope}\\
    \hline
    \textbf{WISDM}   & (36, 6)                        &
    \makecell{Walking/Jogging \\ Sitting/Standing\\ Upstairs/Downstairs} & Accelerometer\\
    \hline
    \textbf{HAR70+}  & (18, 7)                        &
    \makecell{Walking/shuffling \\ Upstairs/Downstairs\\Standing/Sitting/Lying} & Accelerometer\\
    \hline
    \textbf{HARTH}   & (22, 12)                       &
    \makecell{Walking/shuffling/Runing \\ Upstairs/Downstairs\\Standing/Sitting/Lying\\Cycling (sit,stand)\\Cycling inactive (sit,stand)} & Accelerometer\\
    \hline
    \textbf{PAMAP2}  & (8, 18)                        &
    \makecell{Walking/Runing\\Upstairs/Downstairs \\ Standing/Sitting/Lying/Cycling \\ 10 more daily living activities} & IMU \\
    \bottomrule
  \end{tabular}
\end{table*}

\begin{itemize}
\item 
\textit{Human Activity Recognition Using Smartphones (UCI HAR)}~\cite{Anguita2013APD}: This dataset comprises data collected from 30 volunteers aged between 19 and 48, using sensors embedded in smartphones. It includes records of six different activities, such as walking, sitting, standing, going upstairs, going downstairs, and lying down.
\item 
\textit{Wireless Sensor Data Mining (WISDM)}~\cite{kwapisz2011activity}: This dataset employs a methodology simulating real-life activities by placing smartphones in participants' pants pockets and utilizing the sensors therein to collect activity data.
\item 
\textit{HAR70+}~\cite{ustad2023validation,misc_har70+_780}: This dataset is unique as it gathers data from 18 individuals aged between 70 and 95, wearing two accelerometer sensors, documenting eight activity states. The inclusion of elderly subjects holds significant implications for HAR applications, enabling model training for behavior prediction and risk mitigation, such as elderly fall detection~\cite{chen2023digital}.
\item 
\textit{HARTH}~\cite{Logacjov2021HARTHAH}: This dataset comprises recordings from 22 participants wearing two 3-axis Axivity AX3 accelerometers for approximately two hours in free-living environments. One sensor is attached to the right anterior thigh, while the other is affixed to the lower back, with a sampling rate of 50Hz. It captures 12 activity states.
\item 
\textit{PAMAP2}~\cite{Reiss2012IntroducingAN}: The PAMAP2 dataset utilizes three inertial measurement units (IMUs), each containing three-axis MEMS sensors (two accelerometers, one gyroscope, and one magnetometer), all sampled at 100Hz. Nine subjects, comprising eight males and one female aged between 23 and 31, participated in data collection. Notably, in our experiments, due to the minimal data volume for subject 9 after preprocessing, we did not include subject 9's data as a client for FL training.
\end{itemize}

\subsection{Experimental Settings}
In our experiments, we employed the FedAvg algorithm \cite{mcmahan2017communication} to aggregate locally trained models uploaded by clients. We conducted experiments on both the relatively simple  2$\times$ Conv model and the more complex ResNet \cite{he2016deep} model to observe whether model complexity affects the success rate of Membership Inference Attacks (MIA) in our experiments. For the number of clients $k$ included in $D_{mix}$, we conducted experiments with $k=2$ and $k=3$, utilizing the xgboost binary classification model \cite{chen2016xgboost} as the attack model. To distinguish the client $C_j$  targeted by the attacker, we adopted a random selection method and obtained $D_{member}$  by randomly selecting data from $D_j$. Then, when $k=2$, we randomly selected one client ID not equal to $j$, and when $k=3$, we randomly selected two client IDs not equal to $j$, from which we randomly selected data to form $D_{nonmember}$. The data randomly selected here could be used for model training by the clients or for testing purposes, with no requirement regarding their participation in model training. Finally, we partitioned an equal amount of data from $D_{member}$ and $D_{nonmember}$ to form the test dataset $D_{mix}$.

In the adversarial scenario, the attacker selects $C_j$  to participate in the training of the global model in the first round. At this stage, the model uploaded by $C_j$, denoted as $M_{j}^{e}$, has not yet been aggregated. Hence, this model exhibits the poorest generalization to the data of other clients but stronger generalization to its own data. The differences in the distribution of prediction vectors obtained after feeding data into $M_{j}^{e}$  become more pronounced, thereby enhancing the attacker's performance. In subsequent training rounds, the attacker no longer changes the selection of participating clients. However, when $C_j$  is chosen to participate, the attacker still preserves the model uploaded by $C_j$, thereby obtaining more diverse training data for the attack model. Additionally, we conducted experiments on the FedProx algorithm \cite{li2020federated}.

Additionally, existing methodologies commonly employ regularization techniques to mitigate overfitting in machine learning (ML) models and alleviate generalization discrepancies among different client models, thus defending against Membership Inference Attacks (MIA) in federated learning. The purpose of regularization is to reduce the degree of overfitting in the target model, thereby mitigating MIA. Commonly used regularization methods include $L_2$ regularization and dropout. $L_2$ regularization helps prevent excessive model parameters, reducing the model's sensitivity to training data and enhancing its generalization ability. Dropout aids in reducing overfitting in neural networks by compelling the network to learn more robust feature representations, rather than relying excessively on specific neurons. In our study, we focused on investigating the impact of $L_2$ regularization and Dropout defense mechanisms on the attack model.

We primarily select two metrics to assess the discriminative performance of attack models on client data, namely \textit{Accuracy} and \textit{Recall} on the test dataset $D_{mix}$. A higher \textit{Accuracy} value indicates a stronger discriminative capability of the attack model on the data. A higher \textit{Recall} value indicates that the attack model can correctly predict a larger number of data belonging to $C_j$  in the dataset. In the experimental results, we aim to achieve high Accuracy and Recall simultaneously, indicating that the attack model can precisely distinguish whether data belong to $C_j$.

\subsection{HAR Model Performance}
The data in Table \ref{tab:train_test_result} reflect the average accuracy of five HAR datasets on the training and testing sets under different model conditions. These results suggest that after federated learning aggregation, the models exhibit similar accuracy on the training and testing sets, indicating that the models do not overfit the training data but demonstrate good generalization ability. In MIA, attackers can infer whether data participated in training by judging the degree of overfitting of the model to the training data. The higher the degree of overfitting, the better the attack effectiveness. Preliminary experiments indicate that models trained on individual client HAR datasets have very good generalization performance to their own data. For data from the same client, regardless of whether it participated in training, the model's prediction distribution is generally similar. However, there may be differences in the generalization ability of the model to data from different clients. Each client only has good generalization ability to its own data, and the confidence in predicting data from other clients is lower, especially when model training aggregation is not performed. This phenomenon emphasizes the heterogeneity of client data in federated learning and needs to be fully considered during the model aggregation process.

\begin{figure}[t]
  \centering
  \includegraphics[width=0.7\linewidth]{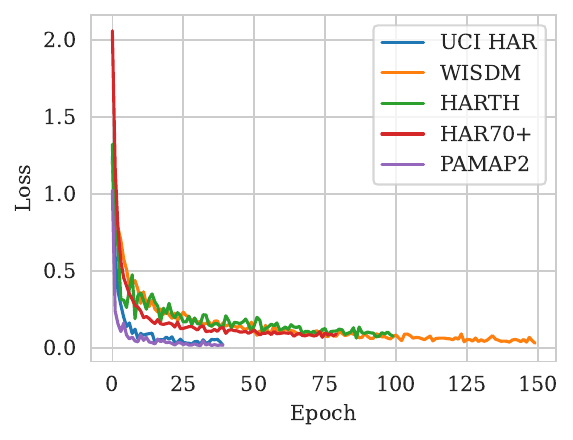}
  \caption{Average loss function curve of the 2$\times$ Conv model during training.}
  \label{fig:cnn_avg_loss}
\end{figure}

\begin{table}[t]
  \centering
\caption{Average training and testing accuracy under various model conditions.}
  \label{tab:train_test_result}
  \begin{tabularx}{0.95\linewidth}{XXXX}
    \toprule
    \multirow{2}{*}{\textbf{Dataset}}         & \multirow{2}{*}{\textbf{Model}} & \multicolumn{2}{c}{\textbf{Accuracy~(\%)}} \\
    \cline{3-4}
    &&\textit{Train}&\textit{Test}\\
    \midrule
    \multirow{3}{*}{UCI HAR} & 2$\times$Conv   & 89.89                       & 88.14                      \\
                             & ResNet         & 91.74                       & 88.38                      \\
                             & FedProx        & 90.16                       & 89.03                      \\
    \hline
    \multirow{3}{*}{WISDM}   & 2$\times$Conv   & 87.15                       & 84.97                      \\
                             & ResNet         & 86.55                       & 84.92                      \\
                             & FedProx        & 84.04                       & 82.72                      \\
    \hline
    \multirow{3}{*}{HAR70+}  & 2$\times$Conv   & 94.48                       & 94.61                      \\
                             & ResNet         & 90.44                       & 90.37                      \\
                             & FedProx        & 93.27                       & 93.27                      \\
    \hline
    \multirow{3}{*}{HARTH}   & 2$\times$Conv   & 90.51                       & 89.58                      \\
                             & ResNet         & 85.24                       & 83.42                      \\
                             & FedProx        & 89.31                       & 88.72                      \\
    \hline
    \multirow{3}{*}{PAMAP2}  & 2$\times$Conv   & 84.25                       & 84.06                      \\
                             & ResNet         & 85.36                       & 85.09                      \\
                             & FedProx        & 85.51                       & 86.05                      \\
    \bottomrule
  \end{tabularx}
\end{table}

\begin{figure*}[t]
  \centering
  \begin{minipage}[t]{0.42\textwidth}
    \includegraphics[width=\textwidth]{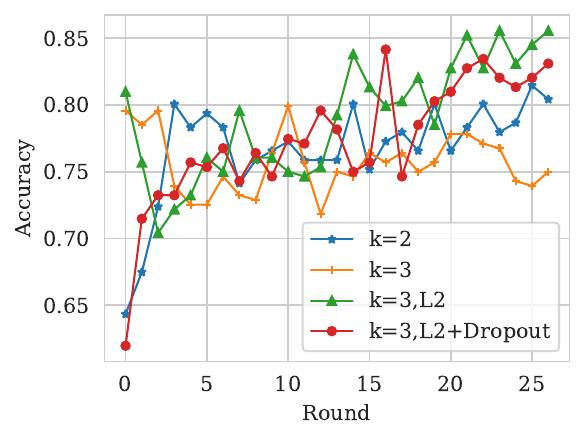}
  \end{minipage}
  \begin{minipage}[t]{0.42\textwidth}
    \includegraphics[width=\textwidth]{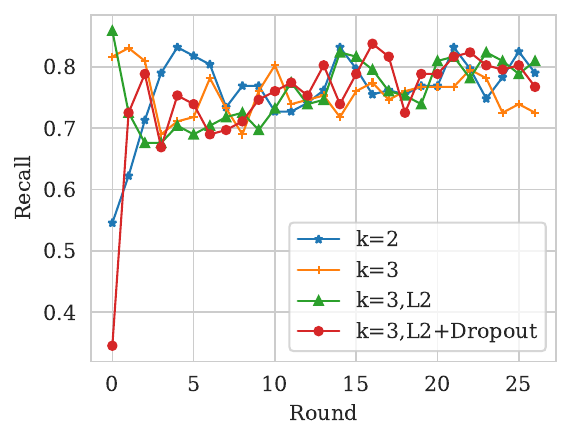}
  \end{minipage}
    \caption{Attack performance of the HAR70+ dataset.}
  \label{fig:har70_accuracy_recall}
\end{figure*}

\begin{table*}[t]
  \centering
    \caption{Attack performance on $D_{mix}$.}
  \label{tab:first_attack_result}
  \begin{tabular}{cccccccccc}
    \toprule
    \multirow{2}{*}{\textbf{Dataset}} & \multirow{2}{*}{\textbf{Model}} & \multicolumn{2}{c}{\textbf{k=2~(\%)}} & \multicolumn{2}{c}{\textbf{k=3~(\%)}} & \multicolumn{2}{c}{\textbf{k=3, $L_2$~(\%)}} & \multicolumn{2}{c}{\textbf{k=3, $L_2$+Dropout~(\%)}} \\
    \cline{3-10}
    &&\textit{Accuracy}&\textit{Recall}&\textit{Accuracy}&\textit{Recall}&\textit{Accuracy}&\textit{Recall}&\textit{Accuracy}&\textit{Recall}\\
    \midrule
    \multirow{2}{*}{UCI HAR}  & 2$\times$Conv  & 86.67 &	93.33 &	63.29 &	69.23 & 64.56 &	66.67 &	56.96 &	35.90 \\
                             & ResNet        & 80.00 &	73.33 &	65.82	& 74.36 &	55.70 &	15.38 &	49.37	& 12.82    \\
    \hline
    \multirow{2}{*}{WISDM}  & 2$\times$Conv  & 82.11 &	86.24 &	81.82 &	87.76 &	89.90 &	91.84 &	67.68	& 51.02 \\
                             & ResNet        & 77.06 &	74.31 &	80.81 &	79.59 &	65.66 &	73.47 &	79.80 & 93.88 \\
    \hline
    \multirow{2}{*}{HAR70+}  & 2$\times$Conv  & 64.34 &	54.55 &	79.58 &	81.69 &	80.99 &	85.92 &	61.97 &	34.51  \\
                             & ResNet        & 63.99 &	66.43 &	59.86 &	56.34 &	72.89 &	81.69 &	74.65 &	69.01  \\
    \hline
    \multirow{2}{*}{HARTH}  & 2$\times$Conv  & 85.64 &	89.50 &	86.91 &	84.21 &	87.96 &	86.32 &	85.34 &	86.32  \\
                             & ResNet        & 70.72 & 72.93 &	76.96 &	80.00 &	90.05 &	86.32	& 79.58	& 81.05  \\
    \hline  
    \multirow{2}{*}{PAMAP2}  & 2$\times$Conv  & 74.87 &	79.22 &	76.00 &	78.78 &	79.81 &	85.74 &	72.09 &	75.04  \\
                             & ResNet        & 62.78 &	61.67 &	69.55 &	73.68 &	68.28 &	76.91 &	71.42	& 76.23 \\  
    \bottomrule
  \end{tabular}
\end{table*}

\begin{table*}[t]
  \centering
  \caption{Attack performance on $D_{mix}$.}
  \label{tab:attack_result}
  \begin{tabular}{cccccccccc}
    \toprule
    \multirow{2}{*}{\textbf{Dataset}} & \multirow{2}{*}{\textbf{Model}} & \multicolumn{2}{c}{\textbf{k=2~(\%)}} & \multicolumn{2}{c}{\textbf{k=3~(\%)}} & \multicolumn{2}{c}{\textbf{k=3, $L_2$~(\%)}} & \multicolumn{2}{c}{\textbf{k=3, $L_2$+Dropout~(\%)}} \\
    \cline{3-10}
    &&\textit{Accuracy}&\textit{Recall}&\textit{Accuracy}&\textit{Recall}&\textit{Accuracy}&\textit{Recall}&\textit{Accuracy}&\textit{Recall}\\
    \midrule
    \multirow{2}{*}{UCI HAR}  & 2$\times$Conv  & 87.78 & 93.33 &	70.89 & 71.79 &	74.68 & 74.36 &	68.35 & 69.23  \\
                             & ResNet        & 85.56 & 82.22 & 69.62 & 74.36 & 65.82 & 79.49 & 78.48 & 82.05 \\
    \hline
    \multirow{2}{*}{WISDM}  & 2$\times$Conv  & 84.86 & 85.32 &	91.92 & 91.84 &	77.78 & 85.71 &	75.76 & 79.59 \\
                             & ResNet        & 76.61 & 74.31 & 84.85 & 85.71 & 83.84 & 87.76 & 89.90 & 89.80 \\
    \hline
    \multirow{2}{*}{HAR70+}  & 2$\times$Conv  & 80.42 & 79.02 & 75.00 & 72.54 & 85.56 & 80.99 & 83.10 & 76.76 \\
                             & ResNet        & 73.08 & 73.43 & 69.72 & 76.76 & 72.89 & 73.24 & 87.68 & 88.03 \\
    \hline
    \multirow{2}{*}{HARTH}  & 2$\times$Conv  & 85.08 & 85.08 &	82.72 & 82.11 &	85.86 & 92.63 &	92.15 & 93.68 \\
                             & ResNet        & 77.90 & 79.56 &	75.39 & 73.68 &	79.58 & 84.21 &	84.29 & 77.89 \\
    \hline  
    \multirow{2}{*}{PAMAP2}  & 2$\times$Conv  & 77.09 & 81.77 & 81.17 & 84.38 & 81.51 & 87.44 & 79.56 & 82.34 \\
                             & ResNet        & 80.41 & 85.86 & 83.55 & 87.44 & 80.32 & 83.70 &	78.54 & 81.32 \\  
    \bottomrule
  \end{tabular}
\end{table*}

\subsection{Attack model performance}
In our experimentation, we employed a strategy to simulate attacks on federated learning systems. Specifically, at the outset, we randomly selected a client as the target for the attacker, denoting its client ID as $j$. The attacker's objective was to accurately differentiate data belonging to the target client $C_j$  from the mixed dataset $D_{mix}$. To achieve this, the attacker utilized known member and non-member data along with the model uploaded by the target client  $C_j$  during training, to train a binary classification model to distinguish data in $D_{mix}$.

Upon the selection of the target client for training, the attacker performed one attack training iteration on the model. Table \ref{tab:first_attack_result} demonstrates that when the attacker controlled the training process and the target client $C_j$ was chosen in the initial training round, the resulting attack model was able to successfully differentiate data belonging to $C_j$  in $D_{mix}$ with high accuracy and recall. Table \ref{tab:attack_result} presents the accuracy and recall of the attack model on $D_{mix}$ after multiple training rounds under different conditions. The results indicate that irrespective of the dataset, the attack model's accuracy consistently exceeded 65\% and recall exceeded 68\%, suggesting that despite defensive measures in federated learning systems, attackers could still achieve moderate success.

In summary, through experimentation on five HAR datasets, we found that both simple 2$\times$ Conv and more complex ResNet models trained by attackers could accurately identify over 
65\% of the data in $D_{mix}$. Even at a minimal level, the attack models could effectively distinguish data. Furthermore, we observed that the impact of adding defensive mechanisms on attacks was relatively limited, with a significant reduction in attack success rate only observed on the UCI HAR dataset. Additionally, as the number of source clients increased, the model could better distinguish data belonging to $C_j$. Further detailed analysis of the experimental results for each dataset will follow to comprehensively understand the findings.

\subsubsection{UCI HAR} 
For the UCI HAR dataset, based on the analysis of the experimental results in Tables \ref{tab:first_attack_result} and \ref{tab:attack_result}, the optimal performance of the attack model trained when the model structure was 2$\times$ Conv was observed when $k=2$. In this scenario, $D_{mix}$ contained data from only two different clients, one of which was the target client. After multiple training iterations, the attack model was able to correctly differentiate over 87\% of the data in $D_{mix}$  and achieved a 93.33\% accuracy in correctly identifying data belonging to the target client. This indicates that the attack model could correctly identify data from the target client with a high probability.

However, when $k=3$, representing data from three different clients in $D_{mix}$, the performance of the attack model decreased by approximately 20\%. This decrease could be attributed to the increasing similarity among the selected client data as the number of clients increased. Nevertheless, even in this scenario, the attack model's recall remained above 71.79\%. Additionally, as shown in Table \ref{tab:first_attack_result}, in experiments with ResNet models incorporating $L_2$ regularization and Dropout defense mechanisms, the accuracy of the attack model in the first training round approached 50\%, with recall less than 20\%. This suggests that the attack model was unable to effectively differentiate the data sources. However, with increasing training rounds, the final results of the attack model also exceeded 65\%.

\subsubsection{WISDM} 
Regarding the WISDM dataset, under different experimental settings, both the accuracy and recall of the attack model remained relatively high. It is noteworthy that the performance of the attack model did not decrease as observed in the UCI HAR dataset when $k$ increased from 2 to 3. In Table \ref{tab:train_test_result}, we observed that the global model trained on the WISDM dataset achieved lower accuracy on test datasets from different clients compared to the training dataset, indicating significant differences among client data, preventing the aggregated model from attaining sufficient accuracy and thereby maintaining the performance of the attack model at a high level.

In experiments with $L_2$ regularization and Dropout defense mechanisms, we observed some interesting phenomena. Performance declined in the 2$\times$ Conv model while it improved in the ResNet model. Particularly noteworthy was the decent performance of the attack model after one training iteration, indicating minimal impact of $L_2$ regularization and Dropout defense on the attack model, especially for the ResNet model. This finding suggests that the effectiveness of defense mechanisms may vary depending on the dataset and model architecture, warranting further research and optimization.

\subsubsection{HAR70+} 
For the HAR70+ dataset, the experimental results presented in Figure \ref{fig:har70_accuracy_recall} indicate that as training rounds increased, the performance of the attack model gradually improved. Initially, in the first training round, the accuracy and recall of the attack model were relatively low, but with increasing training rounds, these performance metrics exhibited an upward trend, indicating that after more training rounds, the attack model could more accurately identify and classify data, thereby improving its overall performance. Furthermore, the results showed a seemingly improved performance of the attack model on this dataset after the addition of $L_2$ regularization and Dropout defense mechanisms. This suggests that the defense mechanisms may not have effectively resisted attacks as expected but might have provided additional information, making it easier for attackers to bypass defenses and enhance performance.

\subsubsection{HARTH} 
In the HARTH dataset, we observed that under ResNet conditions, the average performance of the trained attack model was lower than that of the 2$\times$ Conv model. From the results in Tables \ref{tab:first_attack_result} and \ref{tab:attack_result}, it can be seen that after the first training round, the attack model achieved an accuracy of over 70\% and a recall of over 72\%, with slight performance decline with increasing training rounds, eventually stabilizing.

For the 2$\times$ Conv model, in experiments with only $L_2$ regularization, recall increased from 
86.32\% to 92.63\%. With the addition of $L_2$ regularization and Dropout defense mechanisms, in experiments after multiple training rounds, the accuracy of the attack model increased from 85.34\% to 92.15\%. These results indicate that for both ResNet and 2$\times$ Conv models, the performance of the attack model improved with different defense mechanisms. However, the average performance of the attack model under the ResNet model remained lower than that under the 2$\times$ Conv model, possibly due to the complexity of the ResNet model and its higher resilience to attacks.

\subsubsection{PAMAP2} 
Regarding the PAMAP2 dataset, similar results were observed for attack models trained on ResNet and 2$\times$ Conv models. It was noted that the addition of defense mechanisms had minimal impact on the performance of the attack model in both models. After multiple training rounds, performance improved under various conditions. This similarity may stem from the inherent characteristics of the PAMAP2 dataset, leading to consistent performance of different models on this dataset. Moreover, the minimal impact of defense mechanisms on the attack model suggests a certain robustness of this dataset to current defense strategies.

\subsection{Impact of Defense on the Attack Model}
We conducted focused experiments to observe the impact of $L_2$ regularization and Dropout defense mechanisms on the attack model, and the experimental results are organized as shown in Tables \ref{tab:first_attack_result} and \ref{tab:attack_result}. In experiments conducted on these five datasets, we observed that the effects of $L_2$ regularization and Dropout defense mechanisms on the final performance of the attack model were relatively minor. In fact, the results also indicate that adding defense mechanisms only has a certain impact on the performance of the attack model in the initial few epochs of training. After the attacker uploads the trained model from each target client, our method utilizes the latest uploaded model from the target client to retrain and adjust the attack model, effectively countering the influence of these defense mechanisms.

\section{Related Work}
\label{sec:related}
\textit{Federated HAR: }
Researchers have begun exploring the application of federated learning to train Human Activity Recognition (HAR) models, aiming to address issues such as data silos. Tu et al. \cite{tu2021feddl} introduced a method named FedDL, which learns the similarity between user model weights and dynamically shares these weights to expedite convergence while maintaining high accuracy. In real-life scenarios, individuals often perform the same activities in different ways. Thus, Li et al. \cite{li2021meta} introduced meta-learning into federated learning with Meta-HAR, effectively enhancing the model's personalized performance. Shen et al. \cite{shen2022federated} proposed FedMAT, a federated multi-task attention framework, to address the heterogeneity of HAR user data by extracting features from shared and individual-specific data. Concurrently, Ouyang et al. \cite{ouyang2021clusterfl} argued that previous work overlooked the intrinsic relationships between different user data, resulting in poor performance of federated learning in HAR applications. They proposed ClusterFL, which employs clustering learning based on user similarity and introduces two novel mechanisms to enhance accuracy and reduce communication overhead. Yu et al. \cite{yu2021fedhar} identified four challenges facing HAR in real-world applications: privacy protection, label scarcity, real-time requirements, and heterogeneous patterns. They introduced FedHAR, a semi-supervised personalized federated learning framework, to address these challenges. Li et al. \cite{li2023hierarchical} emphasized the need to consider multiple aspects such as accuracy, fairness, robustness, and scalability in practical HAR scenarios. Thus, they proposed FedCHAR, a personalized federated learning framework with robustness and fairness, and further proposed the scalable and adaptive FedCHAR-DC.

\textit{Membership Inference Attacks: }
Membership Inference Attack (MIA) is a method used to infer whether individuals have participated in the training of a model by determining if their data is included in the model's training dataset~\cite{chen2024EdgeLeakage, scif2022privacy}. Once attackers ascertain that certain data was involved in model training, they can deduce the privacy information of the data based on factors such as the model's usage.
Typically, MIA attacks operate in a centralized machine learning setting where attackers have no knowledge of the target model's structure or the distribution of its training data. In such scenarios, attackers can only gain limited information through black-box access to the target model, including the distribution of training data and black-box queries to the target model \cite{7958568, long2018understanding, Salem2018mlleaks, song2021systematic}. For example, in Machine Learning as a Service (MLaaS) environments, attackers could access predicted outputs of input records to the target model.
Membership inference attacks can also be extended to federated learning, aiming to infer whether a given datum belongs to the model's training data ~\cite{melis2019exploiting, nasr2019comprehensive, wang2023member, chen2024fcl} or to identify which specific client the data belongs to ~\cite{hu2021source, hu2024Source}. 
Recently, membership inference attacks have been applied to the emerging field of model forgetting to assess whether private data has been thoroughly forgotten by new models~\cite{snn2023zhou, annal2023unlearning}.

To the best of our knowledge, there is currently no research investigating the potential leakage of private data in Federated HAR for wearable healthcare devices. We aim to pioneer a MIA-focused data leakage methodology in this emerging area.

\section{Conclusion}
\label{sec:conclusion}
This study focuses on addressing data leakage issues in the process of user behavior recognition using wearable devices. We observe significant variations in HAR data among different users, leading to substantial disparities in the distribution of prediction vectors within federated models, thus posing privacy risks. To mitigate this, we propose a curious server-based membership inference attack method to investigate the privacy risks inherent in HAR model training within federated learning settings. Through experiments conducted on five publicly available HAR datasets, our results indicate that the accuracy of malicious server-based membership inference reaches 92\%. Our research preliminarily confirms the significant privacy leakage risks associated with federated training across multiple wearable devices, offering a valuable perspective for the HAR domain.


\end{document}